# Feynman photon path integral calculations of optical reflection, diffraction, and scattering from conduction electrons

Stephen E. Derenzo
Lawrence Berkeley National Laboratory

## ABSTRACT

This paper describes the use of Feynman photon path integrals to compute the probability of detecting reflected, diffracted, and scattered photons at different points in space after interacting with conduction electrons. Five examples are given: (1) a thin parabolic sheet of conduction electrons (e.g. a metal mirror) that produces a sharp focus of a distant point source surrounded by the Airy diffraction pattern, (2) the loss of focusing power as the thickness of the parabolic sheet is increased and complete destructive interference for thicknesses that are an integer multiple of 1/2 the wavelength, (3) diffraction of photons entering a thin sheet from the side, (4) diffraction of photons entering the side of a sheet as its thickness is increased, and (5) the angular scattering distribution of internally generated photons in an extended volume of conduction electrons. The calculations integrated the complex probability amplitudes for photon paths from (a) a point source to (b) all points in the conduction electron volume and to (c) a point detector. At each detector position the detection probability was computed as the square of the absolute value of the integral. In general, if there is a concentration of paths that have nearly the same complex amplitude phase, reflection dominates. Otherwise, if the conduction electron volume has sharp boundaries, diffraction dominates. Isotropic scattering dominates for conduction electrons distributed throughout an extended volume, and may explain how scintillation photons in cryogenic *n*-type GaAs can escape total internal reflection trapping, which is essential for its high luminosity.

Keywords:
Feynman path integral
conduction electron
parabolic mirror
Airy diffraction
optical diffraction and scattering
dark matter detection

## 1. Introduction

This paper describes the use of Feynman photon path integrals [1-4] to compute the probability of detecting reflected, diffracted, and scattered photons at different points in space after elastic interactions with different volumes containing conduction electrons. It was largely motivated by the discovery that *n*-type GaAs is a bright cryogenic scintillator [5-7] and the supposition that most of the reported narrow-beam absorption [8, 9] is not absolute absorption but a *novel* optical scattering mechanism that allows the scintillation photons to avoid internal reflection trapping [10]. It is an attractive target material for detecting low-energy electronic excitations from interacting dark matter and has no apparent afterglow [5]. It remains metallic and has delocalized conduction electrons even at temperatures close to 0 K for carrier concentrations above the Mott transition at 8 x $10^{15}$/cm$^3$ [5, 11].

In quantum mechanics if an optical photon interacts only elastically with conduction electrons, the probability of its detection at any point in space after reflection, diffraction, or scattering is the square of the absolute value of the integral of the complex amplitudes of all possible paths from (a) the source to (b) all points the conduction electron volume and to (c) the detector. The Feynman lectures [1, 3] describe the use of photon path integrals to explain the optical reflection of a single photon from a flat mirror in a qualitative way with the conclusions that (1) only paths that have very similar lengths and phases can contribute to the reflected photon detection probability, (2) those paths correspond to the classical law of reflection, and (3) the other paths (the vast majority) cancel by destructive interference. This work extends that approach and calculates numerical values of photon path integrals and detection probabilities for various situations, also without relying on classical optics.

The narrow-beam absorption coefficient for 1 μm photons in *n*-type GaAs is approximately proportional to the carrier concentration over two orders of magnitude [8, 9], which makes it possible to compute an interaction cross section. With 4.9 x $10^{17}$ carriers per cm$^3$ the narrow-beam absorption coefficient is 2.5 per cm, which corresponds to an interaction cross section of 5 x $10^{-18}$ cm$^2$. This is almost $10^7$ times larger than the cross section for Thomson scattering on an isolated electron (6.65 x $10^{-25}$ cm$^2$) [12, 13]. In gold the extinction coefficient (imaginary part of the refractive index) is fairly constant between 0.25 μm and 0.50 μm (above interband transitions and below plasma oscillations) and averages 1.9. The corresponding penetration depth is 42 nm. In that thickness there are 2.5 × $10^{17}$ conduction electrons per cm$^2$, which corresponds to a cross section of 4 × $10^{-18}$ cm$^2$. The similarity in cross sections between gold (4 × $10^{-18}$ cm$^2$) and *n*-type GaAs (5 × $10^{-18}$ cm$^2$) arises because in both cases the conduction electrons are delocalized and the positive ions provide charge compensation.

In this work the 1 μm photons are assumed to have an energy below the band gap and only interact elastically with conduction electrons. Photons that pass through the volume of conduction electrons without interacting are ignored. This paper is organized as follows: Section 2. describes the focusing abilities of parabolic sheets of conduction electrons of different thicknesses. Section 3. describes the diffraction of photons entering sheets of conduction electrons of different thicknesses from the side. Section 4. shows that the angular scattering distribution of internally generated photons in an extended volume of conduction electrons is isotropic. Section 5. is a discussion of the results.

## 2. Parabolic sheets of conduction electrons

Fig. 1 shows a circular parabolic sheet of conduction electrons of radius $R$ and focal length $Z_f$. Paths from a distant on-axis point source are parallel and uniformly distributed over the focal plane. Due to the parabolic shape, all paths from the focal plane to a thin ($Z_w = 0$) reflecting surface and back to a point detector at the focal point have the same length, and interfere constructively. It is assumed that direct paths from the source to the detector are blocked. See Section 2.2 for cases where $Z_w > 0$.

The path length $L$ (Eqn. 1) is the sum of (a) the distance from the focal plane to a point $x, y, z$ in the parabolic sheet and (b) the distance from $x, y, z$ to a point detector located at $x_d, y_d, z_d$. For convenience all path lengths are set to zero as they pass through the focal plane.

$$L(x,y,z) = (Z_f - z) + \sqrt{(x_d - x)^2 + (y_d - y)^2 + (z_d - z)^2} \qquad (1)$$

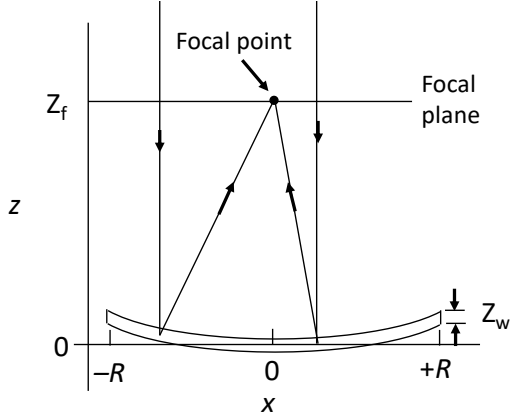

FIG 1. Parabolic sheet of conduction electrons and focal point for on-axis rays from a distant point source. For a thin sheet ($Z_w = 0$) all paths from the distant point source to a point detector at the focal point have the same path length and phase, and interfere constructively. Increasing the sheet thickness $Z_w$ introduces variations in path length and phase, and a reduction in focusing ability. The sheet thickness $Z_w$ is exaggerated in the drawing for clarity.

The $z$ coordinate follows the parabolic surface with curvature parameter $a = 0.25/Z_f$ and sheet displacement $z_0$

$$z = z_0 + a(x^2 + y^2)$$

Each photon path has a complex amplitude phase angle that increases by one rotation ($2\pi$ radians) per wavelength of linear advance. The accumulated phase angle $\theta$ at the detector for path length $L$ and wavelength $\lambda$ is

$$\theta(x,y,z) = 2\pi\ L(x,y,z)/\lambda$$

The complex amplitude $\phi$ for a path with point $x, y, z$ in the conduction electron sheet is:

$$\phi(x,y,z) = e^{i\theta(x,y,z)} = \cos(\theta(x,y,z)) + i\sin(\theta(x,y,z))$$

In quantum mechanics the detection probability $P$ is the square of the absolute value of the integral of the complex amplitudes over the volume of the sheet (Eqns. 2, 3) [1]. $P$ is normalized by the integration volume $A$ and is equal to one when all complex path amplitudes have the maximum possible absolute value of one. One counter-intuitive aspect of quantum mechanics is that the detection probability does not depend on the lengths or propagation times of the photon paths but on the integral of their complex amplitudes at the detector.

$$P(x_d, y_d, z_d) = \frac{1}{A}\left|\int_{-Z_w/2}^{+Z_w/2}\left(\int_{-R}^{+R}\left(\int_{-\sqrt{R^2-y^2}}^{+\sqrt{R^2-y^2}} \phi(x,y,z)dx\right)dy\right)dz_0\right|^2 \quad (2)$$

$$A = \left|\int_{-Z_w/2}^{+Z_w/2}\left(\int_{-R}^{+R}\left(\int_{-\sqrt{R^2-y^2}}^{+\sqrt{R^2-y^2}} dx\right)dy\right)dz_0\right|^2 \quad (3)$$

For $Z_w = 0$ the integral over $z_0$ is ignored.

### 2.1. Focusing and Airy diffraction in the focal plane of a thin parabolic sheet

Highly conductive metals such as silver and gold have a high concentration of delocalized conduction electrons (about $7 \times 10^{22}/cm^3$). A physically thick metal mirror is actually "thin" because incident optical photons can only interact with conduction electrons in the penetration depth, which is much smaller than the wavelength.

Fig. 2 shows the detection probability $P$ as a function of the detector position $x_d$ in the focal plane of a thin ($Z_w = 0$) circular parabolic sheet with radius $R = 5$ mm and focal length $Z_f = 1000$ mm. For each detector position about $10^6$ complex probability amplitudes were computed over the 78.54 mm² surface. The detection probabilities were computed by quadratic interpolation and integration (Eqns. 2,3). At the focal point ($x_d = y_d = 0$, $z_d = Z_f$) all paths contribute constructively with the same phase and $P = 1$. In classical terms, the law of reflection is perfectly maintained over the entire sheet and incident photons from a distant on-axis point source are reflected to meet at the focal point.

The calculated diffraction pattern is in excellent agreement with the classical Airy pattern for a circular aperture. The first four minima appear at 0.122, 0.223, 0.324, and 0.424 mm, the position of the zeros of the first order Bessel function. The calculated detection probabilities at the four minima are below $10^{-13}$, limited by roundoff error. The peaks between the minima have $P$ values of 0.01750, 0.004156, 0.001602, and 0.000779, within 0.1% of their expected values. The total probability density contained in the first minimum ring (i.e. the Airy disk) is 83.8% [14]. Even though each point on the parabolic surface contributes maximally to the detection probability at the focal point, 16.2% of the detection probability lies in the Airy diffraction rings.

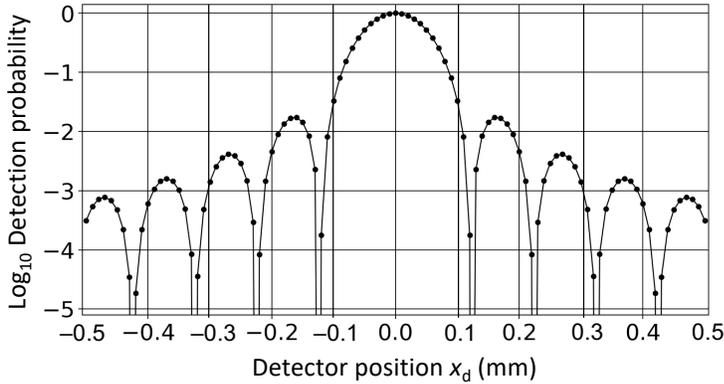

FIG 2. Detection probability for a thin parabolic sheet with $R = 5$ mm, $Z_f = 1000$ mm, and $Z_w = 0$ as a function of detector position $x_d$ in the focal plane. $y_d = 0$, $z_d = Z_f$. See Fig. 1.

## 2.2. Destructive interference in a parabolic sheet as a function of thickness $Z_w$

Figure 3 shows the detection probability $P$ at the focal point vs. the sheet thickness $Z_w$. As $Z_w$ is increased from zero, the resulting variations in phase from different $z$ depths cause phase cancellation and a reduction in $P$. When the sheet thickness is an integer multiple of $\lambda/2$, the integration in $z_0$ is taken over an integer multiple of cycles and the integral is zero. The calculated detection probabilities at the minima were below $10^{-11}$, limited by roundoff error. For a sheet with a thickness of 0.5 μm ($\lambda/2$) the integrations over $x$, $y$, and $z$ used about $10^8$ probability amplitudes for each thickness value.

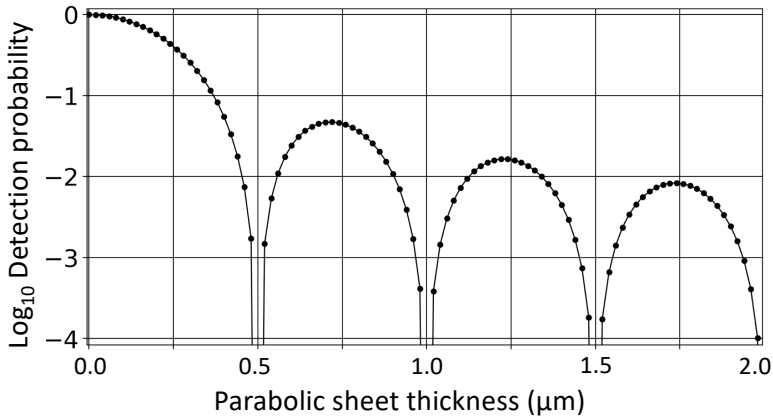

FIG 3. Detection probability at the focal point vs. sheet thickness $Z_w$ from 0 to 2 μm. Destructive interference occurs for $Z_w$ values that are integer multiples of 0.5 μm ($\lambda/2$).

The detection probability for a thick sheet *cannot* be thought of as the average detection probability of a thin sheet during displacement in the $z$ direction. When the thin parabola of Sect. 2.1 is displaced in $z_0$ by 0.25 μm in either direction from $z_0 = 0$, $P$ varies by only a small amount from 1.0 to 0.99999968. In contrast, integrating the complex amplitude over this same range in $z_0$ causes complete destructive interference and $P = 0$. This shows that the complex amplitudes *must* be integrated before squaring.

## 3. Sheets of conduction electrons illuminated from the side

Figure 4 shows a sheet of conduction electrons $D$ wide and $Z_w$ thick illuminated from the side. The detector is at a distance $z_d$ above the plane of the sheet. It is assumed that direct paths from the source to the detector are blocked.

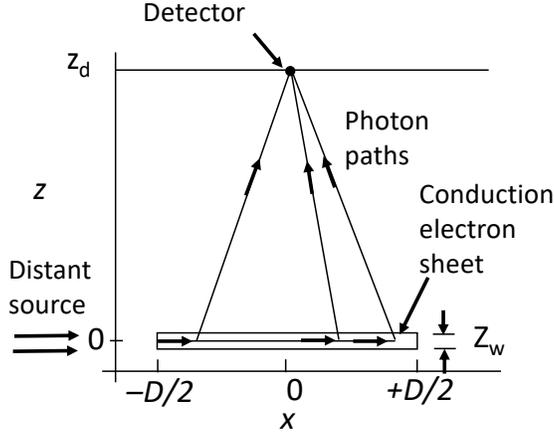

FIG 4. Sheet of conduction electrons of width $D$ and thickness $Z_w$ illuminated from the side.

The photon path length $L(x, z)$ from the left edge of the sheet ($x = -D/2$) to a point $x, z$ in the sheet and to a point detector at $x_d, z_d$ is

$$L(x, z) = x + D/2 + \sqrt{(x_d - x)^2 + (z_d - z)^2}$$

The complex amplitude phase angle at the detector is

$$\theta(x, z) = 2\pi L(x, z)/\lambda$$

The complex amplitude is

$$\phi(x, z) = e^{i\theta(x,z)} = \cos(\theta(x, z)) + i \sin(\theta(x, z))$$

The detection probability $P$ as a function of the detector position $x_d, z_d$ is

$$P(x_d, z_d) = \frac{1}{A}\left|\int_{-Z_w/2}^{+Z_w/2}\left(\int_{-D/2}^{D/2}\phi(x,z)dx\right)dz\right|^2 \tag{5}$$

$$A = \left|\int_{-Z_w/2}^{+Z_w/2}\left(\int_{-D/2}^{D/2}dx\right)dz\right|^2 \tag{5}$$

For $Z_w = 0$ the integral over z is ignored.

### 3.1. Diffraction from a thin sheet

Figure 5 is a plot of the detection probability $P$ as a function of the detector position $x_d$ for $D = 1$ mm, $z_d = 10$ mm, and $Z_w = 0$ (see Fig. 4). $P$ was computed by quadratic interpolation and integration for $10^5$ points along the $x$ axis for each detector position. For the minimum value of $P$ at $x_d = 0$, $L$ increases monotonically from 10.01249 mm at $x = -0.5$ mm to 11.01249 mm at $x = +0.5$ mm and the integral was numerically computed over the 1000 cycles of oscillation. For the peak value of $P$ at $x_d = 5$ µm, $L$ increases from 10.01274 mm at $x = -0.5$ mm to 11.01224 mm at $x = +0.5$ mm and the integral was computed over the 999.5 cycles of oscillation. The minima in Fig. 5 correspond to integrals over an integer number of cycles and the peaks correspond to an integer number plus 0.5 cycles. The sheet acts as a diffuse emitter of width $D$ and the diffraction minima shown in Fig. 5 are spaced $\Delta x_d = 10$ μm apart, in agreement with the classical rectangular aperture formula

$$\sin(\theta_m) = \frac{m\lambda}{D} = m \times 10^{-3}, \text{ and } x_d = z_d \tan(\theta_m) \approx m \times 10 \text{ μm},$$

where $\theta_m$ is the angle of diffraction of the m$^{th}$ minimum.

Because the paths do not have a concentration of similar complex amplitude phases at the detector, the photon is not reflected. Instead, the detection probabilities are small and distributed in angle. The diffraction pattern arises because the conduction electrons are in a volume with sharp boundaries (−$D$/2 and +$D$/2 in this example).

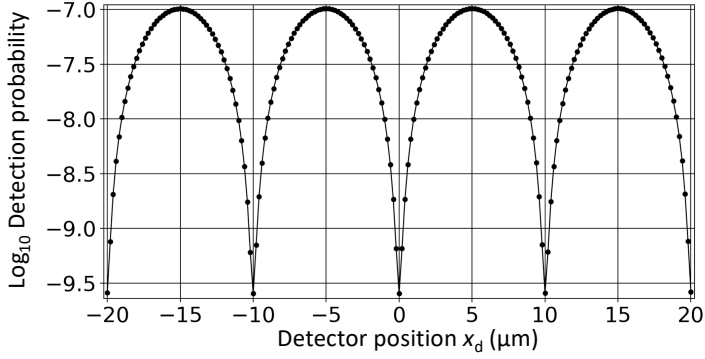

FIG 5. Detection probability as a function of detector position $x_d$ for a thin sheet with $D = 1$ mm, $Z_w = 0$, and $z_d = 10$ mm (see Fig. 4). The spacing of the diffraction minima agree with the classical rectangular aperture formula.

### 3.2. Destructive interference as a function of sheet thickness

Fig. 6 shows the detection probability as a function of sheet thickness $Z_w$ for a sheet with $D = 1$ mm, detector position $z_d = 10$ mm, and $x_d = 5$ μm (a point of maximum detection probability in Fig. 5). When $Z_w$ is a multiple of the wavelength the integration in $z$ is taken over an integer multiple of full cycles and the integral is zero. The calculated detection probabilities at the minima were below $10^{-11}$, limited by roundoff error. For a sheet thickness equal to the wavelength, the integrations over $x$ and $z$ used about $10^7$ probability amplitudes for each thickness value.

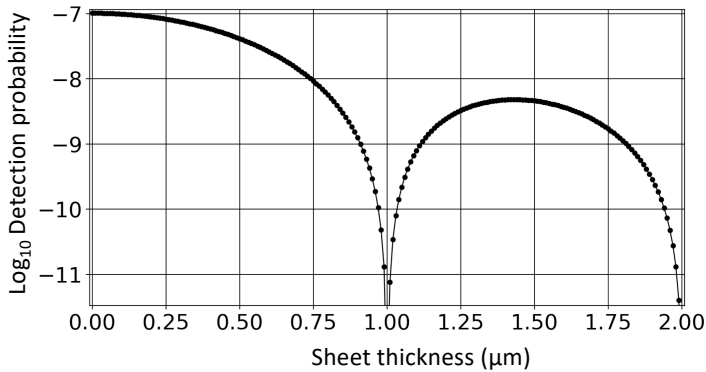

FIG 6. Detection probability as a function of sheet thickness $Z_w$ for $D = 1$ mm, $z_d = 10$ mm, and $x_d = 5$ μm (see Fig. 4).

### 4. Scattering of internally generated photons in an extended volume of conduction electrons

Internal scattering can be modeled by placing the source and detector in a large sphere of conduction electrons. Without loss of generality, the source can be placed at the center of a sphere of radius $R$ and the detector at a point along the $x$-axis at $x = r$, where $R \gg r \gg \lambda$. The photon path length from the source at 0,0,0 to a point $x, y, z$ in the sphere and to the detector at $r, 0, 0$ is

$$L(x,y,z) = \sqrt{x^2 + y^2 + z^2} + \sqrt{(r-x)^2 + y^2 + z^2}$$

The complex amplitude at the detector is

$$\phi(x,y,z) = e^{i2\pi L(x,y,z)/\lambda}$$

The detection probability $P(r)$ is

$$P(r) = \frac{1}{A}\left|\int_z \int_y \int_x \phi(x,y,z)\,dx\,dy\,dz\right|^2$$

$$A = \left|\int_z \int_y \int_x dx\,dy\,dz\right|^2$$

Moving the detector to any other point that is at the same distance $r$ from the center of the sphere is equivalent to a rotation of the coordinate system and will produce the same value of $P$. Therefore the angular distribution for single scattering detected in an extended spherical distribution of conduction electrons is isotropic. The diffraction pattern $P(r)$ can be arbitrarily reduced in scale by making the sphere sufficiently large.

## 5. Discussion

Feynman photon path integrals are an accurate method for computing the focusing ability of a parabolic mirror as well as the Airy diffraction pattern. For a distant on-axis point source each point on the surface of the mirror contributes a maximum complex amplitude to the detection probability at the focal point. However, 16.2% of the detection probability lies in the diffraction pattern, showing that even in the case where every photon is reflected to the focal point, diffracted photons cannot be avoided. When the conduction electrons are distributed throughout a 3-dimensional volume, the law of least time does not apply and diffraction patterns dominate. As the size of the volume is increased the features of the diffraction pattern contract and detection occurs with a low probability over a large range of angles, effectively appearing as isotropic scattering.

The novel optical scattering in $n$-type GaAs has a conduction electron cross section that is almost $10^7$ times larger than Thomson scattering from a single electron ($6.65 \times 10^{-25}$ cm$^2$). It is remarkable that this cross section ($5 \times 10^{-18}$ cm$^2$) does not depend on the conduction electron concentration from $10^{17}$/cm$^3$ to $5 \times 10^{18}$/cm$^3$, and is almost the same as the cross section ($4 \times 10^{-18}$ cm$^2$) for interactions with the conduction electrons ($5.9 \times 10^{22}$/cm$^3$) in a gold mirror. The constant cross section is consistent with the classical picture that an incident photon interacts with a single conduction electron, absorbing and emitting a photon of the same energy without interference from the other conduction electrons. On the other hand, quantum mechanics dictates that the detection probability is proportional to the square of the path integral, which depends on all of the conduction electrons. This contradiction is one of the mysteries of quantum mechanics.

## Acknowledgements

I thank F. Moretti and M. Garcia-Sciveres for helpful discussions and the Lawrence Berkeley National Laboratory for remote access to the literature.